\documentstyle[12pt,epsfig]{article}
\begin{document}

 \title{PT phase transition in higher dimensional quantum systems}

\author{ Bhabani Prasad Mandal\footnote{e-mail address: bhabani.mandal@gmail.com }, 
 Brijesh Kumar Mourya\footnote {e-mail address: brijeshkumarbhu@gmail.com}
 and Rajesh Kumar Yadav\footnote {e-mail address: rajeshastrophysics@gmail.com}}
 \maketitle  
 \begin{center}
 Department of Physics,\\
Banaras Hindu University,\\
Varanasi-221005, INDIA. 
\end{center}

\begin{abstract}
We consider a 2d anisotropic SHO with {\bf ixy} interaction and a 3d SHO in an imaginary magnetic field with $\vec\mu_l$.$\vec B$ interaction to study the $PT$ phase transition analytically in higher dimension.Unbroken $PT$ symmetry
in the first case is complementary to the rotational symmetry of the original Hermitian system .$PT$ phase transition ceases to occur the moment the 2d oscillator becomes isotropic.Transverse magnetic field in the 
other system introduces the anisotropy in the system and the system undergoes PT phase transition depending on the
strength of the magnetic field and frequency of the oscillator.
\end{abstract}

 {\bf Keywords:} \ Non-Hermitian Quantum Mechanics; PT symmetry breaking in Higher dimensions; Anisotropy and PT phase transition;

\newpage

\section{Introduction}
Over the past decade PT symmetric non-Hermitian quantum theories have generated huge excitement\cite{1,2,3}.
It has been shown that PT symmetric non-Hermitian system can have the entire spectrum real
if PT symmetric is unbroken \cite{4,5} and hence a fully consistent quantum theory with unitary time 
evolution can be developed if the associate Hilbert space is equipped with appropriate positive definite inner product  \cite{6,7}. 
This exciting result helps this subject to grow enormously \cite{8}-\cite{20}.
Such a non-Hermitian PT symmetric quantum system generally exhibits a phase transition(more specifically PT breaking transition) that separate
two parametric regions, (i) a region of unbroken PT symmetry in which the entire spectrum is real   
and eigenfunctions of the system respect PT symmetry
(ii) a region of broken PT symmetry in which the whole spectrum(or a part of it) is complex and eigenstates of the system are not eigenstate of PT 
 operator \cite{21}. The study of the PT phase transition has been boosted exponentially due to the fact that
such phase transition and its rich consequence are really observed in  variety of physical systems \cite{22}-\cite{pte3}.
However most of the analytical studies on PT phase transitions are restricted to one dimensional systems.
PT phase transition in higher dimension is studied only by few groups \cite{pha1}-\cite{26}.
Levia et al showed that spontaneous breakdown of PT symmetry can occur for certain non central potentials by considering parity
transformation in terms of spherical polar coordinates \cite{pha4}. Non-Hermitian PT symmetric system in two and three
dimensions have been studied perturbatively in Ref \cite{pha3}. In a very recent work Bender et al \cite{26} have studied a PT phase transition
in higher dimensional quantum systems. They considered  four nontrivial non-Hermitian PT symmetric models in two and three dimensions
to study the PT phase transition. They have shown
in all these models the system passes from PT unbroken phase to broken phase when the non-Hermitian couplings exceed a certain critical value
\cite{26}. However their work is based on perturbation techniques and the PT phase transition is shown numerically.

The aim of this present  manuscript is to take this study of PT phase transition in higher dimension 
by considering two more non-Hermitian but PT symmetric models in higher dimension to get further insight of it .
We show the PT phase transition in our models explicitly and the important point in our study is that our results 
are based on exact analytical calculations. 
We further have realized a possible connection between the symmetry of the original Hermitian Hamiltonian and the PT 
phase transition of non-Hermitian system. In the first example we consider an anisotropic harmonic oscillator in two 
dimensions with a PT symmetric non-Hermitian interaction {\bf $ixy$}. To the best of our knowledge this is the first
time this interaction {\bf $ixy$} is realized as $PT$ symmetric Non-Hermitian interaction.
We analytically calculate the critical value $\lambda_c$ 
of the non-Hermitian coupling which depends on anisotropy of the system. For $\lambda\leq \lambda_c$ the entire spectrum is real and the system is in unbroken PT phase. On the other hand when $\lambda > \lambda_c$ the eigenstates are no longer $PT$ symmetric and as a consequence
we have some complex conjugate pair of eigenvalues indicating the broken PT phase of the system. The most interestingly we observe
that $PT$ phase transition occurs only when the original Hermitian system is anisotropic.  The non-Hermitian 
system of two dimensional isotropic oscillator always remain in $PT$-broken phase.
Thus rotational symmetry of the
original Hermitian system is complementary to the unbroken $PT$ symmetry in this model. 

In the other example we consider an isotropic simple harmonic oscillator in three dimensions with charge $q$
in presence of an imaginary magnetic field along $z$ direction. The introduction of imaginary magnetic field
is motivated by a study by Hatano et al  in which they studied localization and delocalization phase transition in superconductor 
in presence of external imaginary magnetic field \cite{28,29}.
We show that the system undergoes a PT phase transition when the strength of the magnetic field $B\geq B_c$,
the critical magnetic field of the system. Alternatively for a fixed magnetic field PT phase transition occurs when the frequency $\omega$ 
of harmonic oscillator becomes half of the cyclotron frequency $\omega_c$. At the transition point the oscillator do not have any dynamics in $x-y$ plane, it only oscillates in the direction of magnetic field with its original frequency.We would further like to add that the imaginary magnetic field along a preferred direction (z direction) creates the anisotropy in this three dimensional model.

 Now we mention the plan of the paper. In the next section we explain the PT symmetry in two and in three dimension
 and some essential feature of PT phase transition. In section three we present two dimensional anisotropic oscillator with PT symmetric non-Hermitian interaction 
to exhibit a PT phase transition analytically. In section four we discuss three dimensional isotropic
oscillator in presence of imaginary magnetic field. Last section is kept
for discussion and conclusion. 

\section{Parity transformation in higher dimension} 
Parity transformation is an improper Lorentz transformation. In two dimension if we write the parity transformation as
\begin{equation}
 \left(\begin {array}{clcr}
x' \\
y'\\
\end{array} \right) = A\left(\begin {array}{clcr}
x \\
y\\
\end{array} \right) 
\end{equation}
then A is $2\times 2$ real matrix with determinant equal to $-1$.
Now if we consider parity transformation as  simple space reflection in two dimension i.e.
\begin{equation}
x^\prime = -x, \  \ y^\prime = -y  \label{a}
\end{equation}
then the det$ A = +1$, indicating the above transformation(\ref{a}) is a proper Lorentz transformation. In fact the transformation in (\ref{a})
is a rotation of an angle $\pi$ in the $x-y$ plane.This point is explained nicely by Bazeia et al \cite{baz}. The discrete parity transformation in two dimension therefore can be defined
in two alternative ways,
\begin{equation}
P_1 : x^\prime = -x, \ \ y^\prime = y
\end{equation}
\begin{equation}
P_2 : x^\prime = x ,\  \ y^\prime = -y
\end{equation}
Both of these forms $P_1$ and $P_2$ are  equivalent and one can use either of these while checking $PT$ symmetry of a non-Hermitian system in two dimensions. In two dimension parity transformation has also been realized as  \cite{wig}         
\begin{equation}
P_3 : x \longrightarrow  y, \ y \longrightarrow  x
\end{equation}
leading to the transformation matrix
\begin{equation}
 A  =\left( \begin{array}{clcr}
0\ \ 1 \\
1\ \ 0 
\end{array} \right) 
\end{equation}

In fact in the study of non-Hermitian theories, one only requires P to be an involution such that overall $PT$ becomes an antilinear operator. 
 
The structure of parity transformation in three dimension is much more rich.Various different version of anti-linear PT transformation
due the rich structure of parity transformation in 3d has been addressed in detail in Ref. \cite{de}
However in three or in any odd dimension commonly used parity transformation is given as space inversion.
\begin{equation}
x^\prime = -x, \ \  y^\prime = -y,  \ \ z^\prime = -z
\end{equation} 

In even dimension, one has to be careful in defining parity transformation as the determinant of transformation matrix should be -1. 

A $PT$ symmetric
non-Hermitian system is said to be in $PT$ unbroken phase if all the eigenfunctions of the Hamiltonian are also the eigenfunction of the operator $PT$ i.e. $\left[H,PT\right]\psi=0,$ and $\  PT\psi =\pm\psi.$
In this unbroken  $PT$ phase the entire spectrum of the system is real, even through the Hamiltonian is not Hermitian
in the usual sense. Non-Hermitian systems generally depend on certain parameters. Some of the eigenvalues become complex (occur in conjugate pairs) when these parameters change their values. Whether the entire spectrum or only a part of it will be complex depends on the values of the parameter involved in the non-hermitian system. This phase of  the system is described as $PT$ broken phase, even though the non-Hermitian  Hamiltonian
is $PT$ symmetric i.e $[H,PT]\psi =0$, and $\  PT\psi\not=\pm \psi$. Generally system passes from unbroken phase to broken phase when strength of the non-Hermitian coupling increases. Every such system is characterized by a critical value of the coupling below which system is in unbroken phase. In the next two sections we demonstrate $PT$ phase transition in two higher dimensional systems with explicit analytical calculation.

\section{ Anisotropic oscillator in two dimension with Non- Hermitian interaction}

We consider an anisotropic oscillator in two dimension with non- Hermitian interaction described by the Hamiltonian as,

\begin{eqnarray}
H&=&\frac{p_x^2}{2m} +\frac{p_y^2}{2m} +\frac{1}{2}m\omega^2_xx^2 +\frac{1}{2}m\omega^2_yy^2 +i\lambda xy,\qquad \lambda\  \mbox{is  real and }\omega_x\neq\omega_y \nonumber \\ 
\nonumber \\
 &=&  H_0 + H_{nh}
\end{eqnarray}

It is easy to check that this non-Hermitian Hamiltonian is invariant under combined parity\footnote {We can't use $P_3$ as parity transformation 
in 2d for this particular model as the Hermitian part ($H_0$) of this model is not invariant under $P_3$. } and time reversal transformation in two dimension as,

\begin{center}$P_1T(i \lambda xy) =P_2T( i \lambda xy) = i \lambda xy$
\end{center}                           
             
This Hamiltonian can be decoupled by making a coordinate transformation $(x,y)\rightarrow (X,Y)$                                                   
as-
\begin{equation}
H=\frac{P_{X}^2}{2m} +\frac{P_{Y}^2}{2m} +\frac{1}{2} mC_1^2X^2 +\frac{1}{2}mC_2^2Y^2
\end{equation}                
where, 
\begin{equation}  
X=\sqrt\frac{1+k}{2}x -\sqrt\frac{1-k}{2}y ;\qquad     Y=\sqrt\frac{1-k}{2}x +\sqrt\frac{1+k}{2}y ;
\end{equation}
and 
\begin{equation}  
C_1^2 =\frac{1}{2}\left[\omega_+^2- \frac{\omega_-^2}{k}\right] ;\qquad   C_2^2=\frac{1}{2}\left[\omega_+^2 +\frac{\omega_-^2}{k}\right] ; 
\end{equation}                  
where, $\omega_+^2\equiv \omega_x^2 + \omega_y^2 ;\qquad  \omega_-^2\equiv \omega_y^2 -\omega_x^2 ;\qquad k^{-1}  =\sqrt{1-\frac{4\lambda^2}{m^2\omega^4_-}} $ ;\\
The energy eigenvalues and eigenfunctions are written as
\begin{equation} 
E_{n_1,n_2} =(n_1+\frac{1}{2}) \hbar C_1 +(n_2+ \frac{1}{2}) \hbar C_2 
\end{equation}
\begin{equation}
\psi_{n_1, n_2}(X,Y) =N\exp\left[- {(\frac{\alpha_1^2X^2}{2}+\frac{\alpha_2^2Y^2}{2}})\right]  H_{n_1}(\alpha_1X)H_{n_2}(\alpha_2Y) 
\end{equation}
where,   \  $\alpha_1^2=\frac{mC_1}{\hbar }$  and  $\alpha_2^2 =\frac{mC_2}{\hbar }$.\\
Now we consider the case when
$k$ is real i.e. $|\lambda|\le|\frac{m\omega_-^2}{2}|$.
 In this case,
\begin{equation} 
C_1^2 = \frac{1}{2}[\omega_x^2(1 +\frac{1}{k}) +\omega_y^2(1 -\frac{1}{k})]> 0,\qquad
C_2^2 = \frac{1}{2}[\omega_x^2(1 -\frac{1}{k}) +\omega_y^2(1 +\frac{1}{k})] >  0, 
\end{equation}
as $k\geq 1$. This further leads to entire real spectrum . The wave function re-expressed in term of (x,y) as             
\begin{eqnarray}                  
\psi_{n_1,n_2}(x,y) &=& N\exp\left[- { \frac{m}{2\hbar }[(C_1+C_2)(x^2+y^2) +(C_2-C_1)2i\lambda kxy}\right]\nonumber\\
&& H_{n_1}[\alpha_1(\sqrt\frac{k+1}{2}x  -i\sqrt\frac{k-1}{2}y)] 
H_{n_2}[\alpha_2(\sqrt\frac{k-1}{2}x +i\sqrt\frac{k+1}{2}y)]
\end{eqnarray}

Under $PT$ transformation
\begin{equation}
P_1 T \psi_{n_1,n_2}(x,y) = (-1)^{n_1}\psi_{n_1,n_2}(x,y) =\pm\psi_{n_1,n_2}(x,y),
\end{equation}
\begin{equation}
P_2 T \psi_{n_1,n_2}(x,y) = (-1)^{n_2}\psi_{n_1,n_2}(x,y) =\pm\psi_{n_1,n_2}(x,y)
\end{equation}
as $n_1 ,n_2$ are zero or positive integers.
Therefore $PT$ symmetry is unbroken as long as $|\lambda|\le|\frac{m\omega_-^2}{2}|$ and as a consequence of the entire spectrum is real.
It is straight forward to check for $|\lambda|>|\frac{m\omega_-^2}{2}|$, $k$ is imaginary and hence
$P_iT\psi_{n_1,n_2}(x,y) \not = \psi_{n_1,n_2}(x,y)$ for $i=1,2$
and the spectrum is no longer real. Some of the eigenvalues occur in complex conjugate pairs,the spectrum in this 
situation is written as
\begin{equation} 
E_{n_1,n_2} =(n_1+\frac{1}{2}) \hbar (A-iB) +(n_2+ \frac{1}{2}) \hbar (A+iB) 
\end{equation}
Where A and B are real and can be given as
\begin{equation}  
A^2 =\frac{1}{2}\left[\omega_+^2+ \sqrt{\frac{\omega_+^4 +\omega_-^4}{k_1^2}}\right];\qquad   B^2= \frac{1}{2}\left[- \omega_+^2+ \sqrt{\frac{\omega_+^4 +\omega_-^4}{k_1^2}}\right]
\end{equation}              
$k_1(\equiv-ik)$ is also real.It is clear from equation (18) that $E_{n_1,n_2}$ and $E_{n_2,n_1}$ are
complex conjugate to each other. 
On the other hand energy eigenvalue are real when $n_1=n_2$.
The critical value of the coupling $\lambda_c (=\frac{m\omega^2_- }{2})$ depends on the anisotropic of the system.
If the system is more anisotropic the span of the $PT$ unbroken phase is longer. When the system becomes isotropic,
i.e. $\omega_x = \omega_y$, i.e. $\lambda_c=0$, the system will always lie in the broken PT phase and it will not be possible 
to have entire spectrum real for any condition on the parameters. This result leads to a very important realization for this particular 
systems. Rotational symmetry which leads to isotropy of original Hermitian system $H_0$ is complementary to unbroken PT symmetry of the $PT$ symmetric non-Hermitian system $H$. As long as the original system $H_0$ has rotational invariance, $H$ can not have unbroken PT symmetry. The moment rotational symmetry of $H_0$ breaks the non-Hermitian system becomes capable of going through a PT phase transition.

\section{Three-dimensional isotropic harmonic oscillator  in an external imaginary magnetic field}
Three dimensional isotropic simple harmonic oscillator in external imaginary magnetic field (iB) with $i\mu_l.B$ coupling can be
written as 

\begin{equation}
H =\frac{1}{2m}(\vec{p} -\frac{iq\vec A}{c})^2 + \frac{1}{2}mw^2(x^2 +y^2 +z^2) +i\vec \mu_l.\vec B 
\end{equation}
where q is the charge of the oscillator and $\vec B =\vec \nabla  \times\vec A $.

Assuming the imaginary magnetic field along $z$ direction and considering the vector potential $\vec{A}= \{ -\frac{By}{2},\frac{Bx}{2},0\}$ in 
symmetric gauge it can written as
\begin{equation}
H =\frac{1}{2m}(p_x +\frac{iqy B}{2c})^2 +\frac{1}{2m}(p_y -\frac{iqx B}{2c})^2 +\frac{1}{2m}(p_z)^2 +\frac{1}{2}m\omega^2(x^2 +y^2 +z^2) +i\mu_{lz} B
\end{equation}
This non-Hermitian Hamiltonian is $PT$ invariant as  parity is considered to be simple space reflection in odd dimension. 
The Hamiltonian in equation (21) further reduced to a 3d anisotropic oscillator
\begin{equation}
H =\frac{p^2_x}{2m} +\frac{p^2_y}{2m} +\frac{p^2_z}{2m} +\frac{1}{2}m\omega^2_1(x^2 + y^2) +\frac{1}{2}m\omega^2 z^2
\end{equation}
\begin{equation}
\mbox{ where}, \omega^2_1 = \omega^2 -\frac{q^2B^2}{4m^2c^2} = \omega^2 -\frac{\omega^2_c}{4}
\end{equation}
$ \omega_c = \frac{qB}{mc}$ is usual cyclotron frequency.
The Schrodinger equation for this system can be solved explicitly and 
 the energy eigenvalue and eigenfunction for this systems are given as,
\begin{equation}
E_{n_x n_y n_z} =(n_x +n_y + 1)\hbar \omega_1 +(n_z +\frac{1}{2})\hbar \omega
\end{equation}
\begin{equation}
\psi_{n_x n_y n_z}(x,y,z) = \exp\left[-\frac{\alpha^2}{2}(x^2 + y^2) +\frac{\alpha^2_1z^2}{2}\right] H_{n_x}(\alpha x)H_{n_y}(\alpha y) H_{n_z}(\alpha_1 z)
\end{equation}

where,
$\alpha^2=\frac {m\omega_1}{\hbar } $,
\  $\alpha^2_1=\frac {m\omega}{\hbar }$.  
If the magnetic field is sufficiently weak, $B\leq \frac{2 m \omega c}{q}$ or for a fixed magnetic field oscillator 
frequency $\omega\geq\frac {\omega_c}{2}$, 
then $\omega_1$ is real and, hence the entire spectrum is real. In this case it is straight forward to check 
$PT \psi_{n_x n_y n_z}(x,y,z) =(-1)^{n_x +n_y +n_z}\psi_{n_x n_y n_z}(x,y,z)$, indicating the system is in unbroken 
$PT$ phase. However if the strength of the magnetic field exceeds a critical value $B>B_c =\frac{2 m \omega c}{q}$ 
or the oscillator frequency is less then half of the cyclotron frequency for fixed magnetic field ie $\omega\leq \frac{\omega_c} {2}$, 
then $\omega_1=\pm \sqrt{\omega^2 -\frac{q^2B^2}{4m^2c^2}}$, becomes complex i.e  $\omega_1\equiv \pm i \tilde\omega$ where $\tilde \omega \equiv \sqrt{-\omega^2 +\frac{q^2B^2}{4m^2c^2}}$ is real. We have pairs of complex conjugate eigenvalues given as
\begin{equation}
E_{n_x n_y n_z} =\pm i(n_x +n_y + 1)\hbar \tilde \omega +(n_z +\frac{1}{2})\hbar \omega
\end{equation}
and system is in broken phase of $PT$ ,this is because $\omega_1$ is imaginary and under PT it changes to $-\omega_1$ 
hence $PT \psi_{n_x n_y n_z}(x,y,z) \pm\psi_{n_x n_y n_z}(x,y,z)$. The system is PT unbroken phase if $ 0<B\le B_c$ 
for fixed oscillator frequency $\omega$
and $PT$ phase transition occurs as the strength of magnetic field exceeds the critical value.            for a fixed 
magnetic field ,the system is in unbroken $PT$ phase if oscillator frequency is grater or equal to half of the cyclotron
frequency $(\omega_c)$

\section {Conclusion}

We have demonstrated the $PT$ phase transition  in higher dimension, with the help of two simple but non-Hermitian
$PT$ symmetric systems. We have considered an anisotropic simple harmonic oscillator in 2d with a  $PT$
symmetric non-Hermitian interaction to show that system undergoes a $PT$ phase transition as long as it is anisotropic.
The critical value of the coupling vanishes when system is isotropic and phase transition ceases to occur.The system remains in the broken phase all time. This leads to an important connection between the rotational symmetry of original Hermitian system (2d SHO) with the unbroken PT symmetry of non-Hermitian systems (2d SHO) with non-Hermitian interaction).They appear complementary to each other. To have a better understanding of this connection one needs to explore it further in other non-Hermitian systems. 
It will be  possible to experimentally observe the $PT$ phase transition in this  system with the help of simple
mechanical experiment similar to one reported in \cite{ben}-\cite{pte3}.
In the second example we consider a charged isotropic oscillator in $3d$ in the influence of an external imaginary magnetic field along $Z$-direction. If  the strength of magnetic field is strong enough such that cyclotron frequency is greater than ${2} $ times the natural frequency of this isotropic oscillator, then the system passes from unbroken phase to broken phase. On the other hand for sufficiently weak magnetic field system remains in $PT$ unbroken phase.It is worth mentioning that magnetic field introduces the anisotropy in this system.It will be exciting to explore the connection of $PT$ phase transition in this model to localization/delocalization phase transition\cite{28}-\cite{29} in superconductor in presence of imaginary magnetic field.

{\bf Acknowledgment}

One of us (RKY) acknowledges the financial support from UGC under FIP Scheme.


\begin{thebibliography}{99}
\bibitem{1} C. M. Bender, S. Boettcher: {\it Phys. Rev.Lett} \textbf {80} 5243 (1998).
\bibitem{2} C. M. Bender:  {\it Rep. Prog. Phys.} \textbf {70947} (2007) and references therein.
\bibitem{3} A. Mostafazadeh: {\it Int.J.Geom.Math.Mod.Phys.} \textbf {7} 1191 (2010) and references therein.
\bibitem{4} C. M. Bender, S. Boettcher: {\it Phys. Rev.Lett} \textbf {89}, 270401-1 (1998) .
\bibitem{5} A. Khare and B. P. Mandal: {\it Phys.Lett} \textbf {A 272} 53 (2000).
\bibitem{6} C. M. Bender, K. Besseghir, H. F Jones and X. Yin:
 {\it J. Phys. A }, \textbf {42} 355301 (2009); Carl M. Bender, Barnabas Tan, {\it J.Phys.A} \textbf 39 1945 (2006); Carl M. Bender, Hugh F. Jones, {\it Phys. Lett. A} \textbf {328} 102 (2004).  
\bibitem{7} A. Mostafazadeh and F. Zamani, Ann. Phys., 321,(2006) 2183 ; ibid (2006).
2210.

\bibitem{8}  A. Mostafazadeh and H. Mehr-Dehnavi : {\it J.Phys} \textbf {A 42} 125303 (2009);\\
            A. Mostafazadeh :{\it Phys. Rev. Lett.} \textbf {102} 220402 (2009);\\
            A.Mostafazadeh :{\it J.Phys.} \textbf A :Math. Theor. 44 375302 (2009);
\bibitem{9}  Z. Ahmed : Phys. Lett. \textbf {A}282 343; 287 295 (2001).            
\bibitem{10}  Z. Ahmed: {\it J. Phys.} \textbf A: Math. Theor. 42 473005 (2001).
\bibitem{11}  A. Khare and B. P. Mandal: Spl issue of {\it Pramana J of Physics} \textbf {73},387 (2009).
\bibitem{12}  B. Basu-Mallick: {\it Int.Jof Mod.Phys.} \textbf{B 16},1875(2002);\\
            B. Basu-mallick, T. Bhattacharyya, A. Kundu and B. P. Mandal :{\it Czech.J.Phys} \textbf{54},\\5(2004);
\bibitem{13}  B. Basu-Mallick, and B. P. Mandal: Phys. Lett.\textbf {A 284},231(2001);\\ B. Basu-mallick, T. Bhattacharyya and
            B. P. Mandal: {\it Mod. Phys. Lett.} \textbf {A 20}, 543(2004);
\bibitem{14}  B. P. Mandal, A. Ghatak, .{\it J. Phys. A}, Math. Gen \textbf {45} 444022 (2012).
\bibitem{15}  B. P. Mandal, S. Gupta: {\it Mod. Phys. Lett.} \textbf {A 25} 1723 (2010).
\bibitem{16}  B. P. Mandal: Mod. Phys. Lett \textbf {A 20} 655(2005).
\bibitem{17}  A. Ghatak, B. P. Mandal: {\it J. Phys. A}, Math. Gen \textbf {45} 355301 (2012).
\bibitem{18}  B. Bagchi, C. Quesne: {\it Phys. Lett.} \textbf {A 273},256(2000).
\bibitem{19}  K. Abhinav, P. K. Panigrahi:{\it Annl. Phys.}  \textbf {326} 538(2011).
\bibitem{20}  A.Ghatak , Z. A. Nathan, B. P. Mandal and Z. Ahmed: {\it J. Phys. A}, Math. Gen \textbf {45} 465305 (2012).

\bibitem{21}  Z. H. Musslimani, K. G. Makris, R. El-Ganainy, D. N. Christodulides: {\it Phys. Rev. Lett.} \textbf{100},030402 (2008).
\bibitem{22}  C. E. Ruter, K. G. Makris, R. El-Ganainy, D. N. Christodulides, M. Segev, and D. Kip: {\it Nature (London)phys.}\textbf{6}, 192(2010).
\bibitem{23}  A. Guo, G. J. Salamo: {\it Phys.Rev.Lett.}\textbf{103} , 093902 (2009).

\bibitem{24} C. M. Bender, S. Boettcher  and P.N. Meisinger : {\it J. Math. Phys.}\textbf{40} 2201 (1999).

\bibitem{25}  C. T. West, T. Kottos, and T.Prosen: {\it Phys. Rev. Lett} \textbf {104}, 054102(2010).

\bibitem{ben} Carl. M. Bender, Bjorn K. Berntson, David Parker and E. Samuel:  arXiv: math-ph/1206.4972v1. 
%
\bibitem{pte1}   Joseph Schindler, Ang Li,Mei C.Zheng,F.M.Ellis,Tsampikos Kottos {\it Phys.Rev.} \textbf {A  84}, 040101(2011).
\bibitem{pte2}   Hamidreza Ramezani,J. Schindler,F.M.Ellis,Tsampikos kottos {\it Phys.Rev.} \textbf {A  85}, 062122(2012).
\bibitem{pte3}   Zin Lin,J. Schindler,Fred.M.Ellis,Tsampikos kottos {\it Phys.Rev.} \textbf {A  85}, 050101(R)(2012). 
%
\bibitem{pha1}   C. M. Bender, G.V.Dunne,P.N.Meisinger,M.Simsek, {\it Phys.Lett} \textbf {A} 281 (2001)311-316.
\bibitem{pha2}   C. M. Bender,M.Berryand A.Mandilara {\it J.Phys} \textbf {35} (2002)L467.
\bibitem{pha3}   M. Znojil, {\it J.Phys} \textbf {A 36} (2003)7825.
\bibitem{pha4}   G. Levai, {\it J.Phys} \textbf {A 41} (2008)244015.
%
\bibitem{26}  Carl M. Bender, David J. Weir : arXiv: quant-phys/1206.5100v1. 
\bibitem{27}  A. Nanayakkara: {\it Phys. Lett.} {\bf A 304}, 67 (2002). 
\bibitem{28}  N. Hatano and D.R.Nelson: {\it Phys.Rev.} \textbf {B  58}, 8384(1998).
\bibitem{29}  N. Hatano and D.R.Nelson: {\it Phys.Rev.Lett.}\textbf {77}, 570(1996).
\bibitem{baz} D. Bazeia, Ashok Das and L. Losano: {\it Phys.Lett.} \textbf {B 673}, 283(2009).
 \bibitem{wig} E.Wigner,J.Math.Phys.1(1960)409
\bibitem{de}   S.Dey,A.Fringe and L.Gouba, {\it J.Phys} \textbf {A 45} (2012)385302.


\end{thebibliography}
\end{document}